\documentclass[10pt,conference]{IEEEtran}

\usepackage{url}

\usepackage{fancybox}

\usepackage{cite}
\usepackage{booktabs}
\usepackage{tcolorbox}
\usepackage{xspace}
\usepackage{balance}

\ifCLASSINFOpdf
\else
\fi
\usepackage{enumitem}

\usepackage{algorithmic}

\usepackage{array}
\newcommand{\reftype}[1]{{\it #1}}

\newcommand{\etal}{et~al.\xspace}

\begin{document}
%
\title{Do Developers Refactor Data Access Code? An Empirical Study}



%

\author{\IEEEauthorblockN{Biruk Asmare Muse}
\IEEEauthorblockA{
Polytechnique Montréal\\
biruk-asmare.muse@polymtl.ca}
\and
\IEEEauthorblockN{Foutse Khomh}
\IEEEauthorblockA{Polytechnique Montréal\\
foutse.khomh@polymtl.ca}
\and
\IEEEauthorblockN{Giuliano Antoniol}
\IEEEauthorblockA{Polytechnique Montréal\\
antoniol@ieee.org}}


\maketitle

\begin{abstract}

Developers often refactor code to improve the maintainability and comprehension of the software. There are many studies on refactoring activities in traditional software systems. However, refactoring in data-intensive systems is not well explored. Understanding the refactoring practices of developers is important to develop efficient tool support.We conducted a longitudinal study of refactoring activities in data access classes using 12 data-intensive subject systems. We investigated the prevalence and evolution of refactorings and the association of refactorings with data access smells. We also conducted a manual analysis of over 378 samples of data access refactoring instances to identify the functionalities of the code that are targeted by such refactorings. Our results show that (1) data access refactorings are prevalent and different in type. \textit{Rename variable} is the most prevalent data access refactoring. (2) The prevalence and type of refactorings vary as systems evolve in time. (3) Most data access refactorings target codes that implement data fetching and insertion. (4) Data access refactorings do not generally touch SQL queries. Overall, the results show that data access refactorings focus on improving the code quality but not the underlying data access operations. Hence, more work is needed from the research community on providing awareness and support to practitioners on the benefits of addressing data access smells with refactorings.

Keywords- refactoring; data-intensive systems; data access classes; database access; empirical study
\end{abstract}


%
\IEEEpeerreviewmaketitle

\section{Introduction}


The vast amount of data produced by digital devices and humans have contributed to the development of data-driven applications, and they are becoming critical for modern civilization; affecting people in all areas of life\cite{Park2021}. 

The development of data-intensive systems typically requires the integration of a multitude of specialized frameworks for data storage (e.g., relational or NoSQL databases), processing (e.g., Hadoop, Spark), and learning (e.g., TensorFlow, Scikit-learn) which poses several 
design, implementation, and quality assurance challenges\cite{Foidl2019,Hummel2018,Park2021}. Developers of data-intensive systems also often face the usual challenges of release pressures which forces them to compromise software quality; introducing technical debt\cite{Foidl2019} and code smells\cite{muse2020prevalence}. Data-intensive systems devote their main functionality to data access and manipulation. Hence, they are prone to both traditional code smells and data access specific code smells (e.g., SQL code smells\cite{nagy2017static}) \cite{muse2020prevalence}. Code smells can be removed by developers through refactoring \cite{fowler2018refactoring}.

Although several researchers have investigated the prevalence (e.g., \cite{vassallo2019large, peruma2019preliminary}), co-occurrence \cite{iammarino2019self}, motivation\cite{peruma2019preliminary,silva2016we} and impacts \cite{chavez2017does,ferreira2018buggy,mahmoudi2019refactorings} of refactoring in traditional software systems, refactoring in data-intensive systems is not yet investigated. The critical nature of data access code in such systems and the potential susceptibility of such systems to both traditional and data access specific smells calls for separate attention to data access refactoring. We define data access refactoring as refactoring performed on data access classes. Data access classes are classes that perform direct interactions with databases or other persistence systems via calls to driver functions or APIs. Data access classes are responsible for implementing data access logic in data-intensive systems. We refer to non data access classes as regular classes and their refactoring as regular refactoring to differentiate from data access classes. Refactoring in data access classes introduces further complexity as it involves read and writes interaction with a database. Developers must take into account improving 
the performance and robustness of data access operations besides improving the understandability and maintainability of the source code. 
Hence separate attention should be given to refactoring data access classes. The goal of this study is to address this gap in the literature by investigating the characteristics of refactoring activities in data access classes. We conduct a case study on open source data-intensive systems to answer the following research questions.

\begin{enumerate}
    \item [RQ1:] \textbf{How prevalent are refactorings in data access classes?}
    We found that refactorings are prevalent in both data access classes and regular classes. Specifically, \textit{Rename variable} is the most prevalent refactoring in data access classes while \textit{Move attribute} is the most prevalent in regular classes. 
    
    \item [RQ2:] \textbf{How  do  refactoring  activities  change  during  the lifetime of the subject systems?}
    We found that refactoring activities are distributed by type and in time as systems evolve. Furthermore, data access refactoring has a tendency to be applied at later stages of the systems' evolution compared to regular refactoring.  
    \item [RQ3:] \textbf{What do code elements targeted by data access refactorings implement?}
    We identified 36 different functionalities that focus on data access, database management, testing, query manipulation, and other functionalities not specific to data access classes such as initialization of components and other helper functions. The most prevalent functionality is data fetching spanning 27.6\% of the analyzed sample refactorings.
    
    \item [RQ4:] \textbf{Do  data access  refactoring  activities  touch  SQL queries and SQL code smells?}
    We found that data access refactorings do not generally touch SQL queries and SQL smells. Using  line level  matching,  only  0.45\% of  refactoring  instances  touched SQL  queries. Furthermore, none of the matched SQL queries had SQL code smell. 
    
\end{enumerate}

\section{Study Design}

\begin{figure}[!ht]
\centering
        \includegraphics[width=\linewidth]{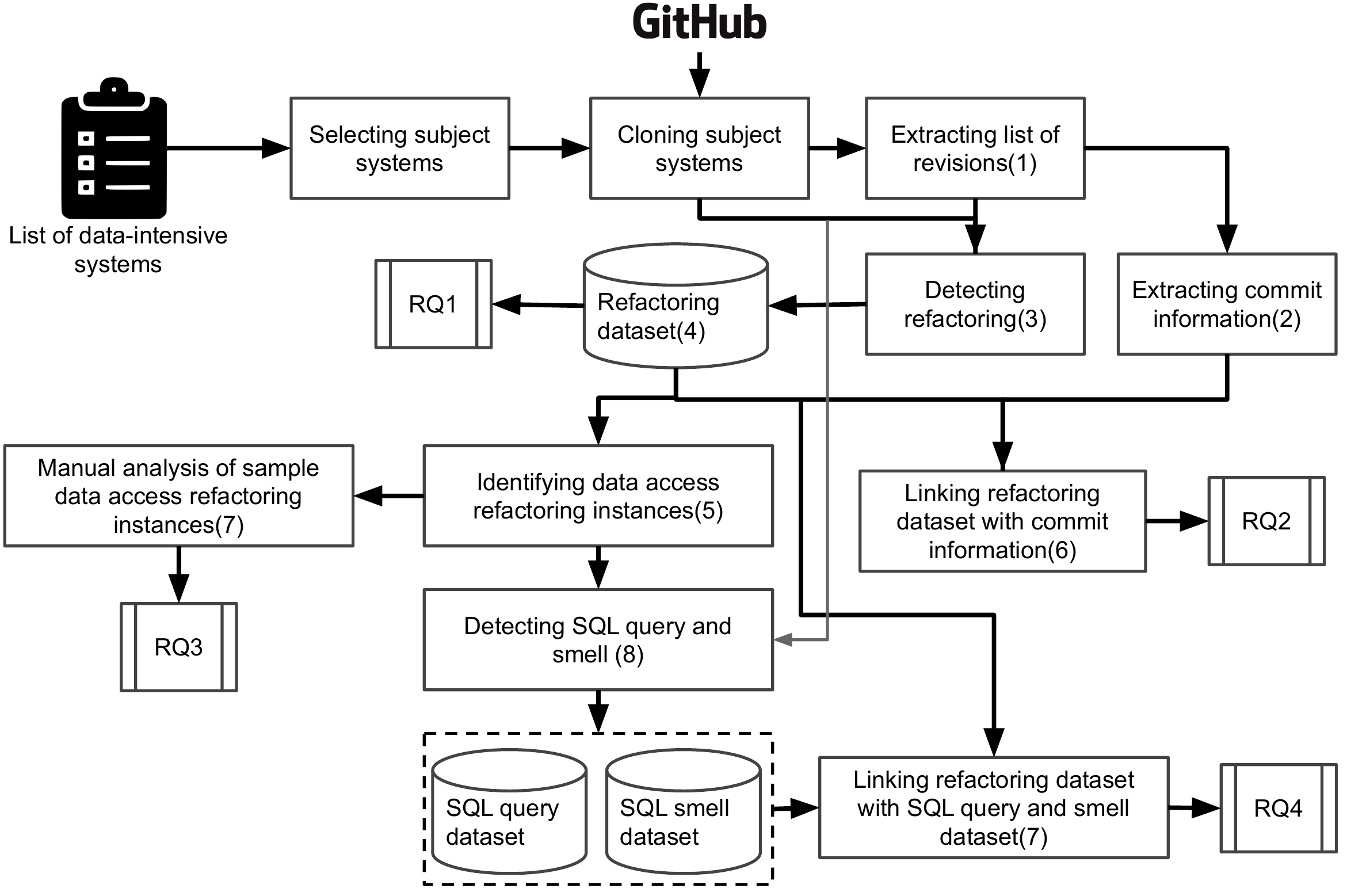}
     
    \caption{Overview of the study method}
    \label{fig:method}
\end{figure}
In this section, we describe the design of our case study that aims to understand refactoring activities in 
data-intensive systems. We address the following four research questions:

\begin{enumerate}[labelindent=\parindent,leftmargin=4\parindent]
    \item [RQ1:] How prevalent are refactorings in data access classes?
    \item [RQ2:] How  do  refactoring  activities  change  during  the lifetime of the subject systems?
    \item [RQ3:] What do code elements targeted by data access refactorings implement?
    \item [RQ4:] Do  data access  refactoring  activities  touch  SQL queries and SQL code smells?
\end{enumerate}


\subsection{Data collection}
The context of this study is data-intensive systems. To select our subject systems, we started with the subject systems identified in the work of Muse \etal \cite{muse2020prevalence} in which they studied the prevalence and impact of SQL smells using data-intensive systems as subject systems. They studied 150 open source data-intensive systems collected from GitHub. Our research questions require analyzing 
every version of the subject systems. Hence, we could not use all of the subject systems due to time and resource constraints. Muse \etal provided the list of the subject systems together with information about the number of SQL queries and number of data access classes per subject system \footnote{https://bit.ly/39RXcPI}. Using  the number of SQL queries as a proxy to pick the most data-intensive systems, we ranked the systems in decreasing order of the number of SQL queries and took the first 12 subject systems. Although \textit{bio2rdf-scripts\footnote{https://github.com/bio2rdf/bio2rdf-scripts
}} project has the highest number of queries, the number of refactoring instances detected was only 6. Hence we removed this project from our list of subject systems. Table \ref{tbl:subjectsystems} shows the summary of our subject systems. We analyzed 2, 473, 090 refactoring instances from 174776 commits. The systems have an average of 519 queries and 67 data access classes. Furthermore, the subject systems have 206, 091 refactoring instances on average.


\begin{table}[]

\centering
\caption{List of subject systems with a number of commits, number of queries, and number of data access files and number of refactoring instances}
\label{tbl:subjectsystems}
\resizebox{0.9\linewidth}{!}{%
\begin{tabular}{@{}lllll@{}}

\toprule
\textbf{Project Name} & \textbf{\begin{tabular}[c]{@{}l@{}}Number of \\ commits\end{tabular}} & \textbf{\begin{tabular}[c]{@{}l@{}}Number of \\ Queries\end{tabular}} & \textbf{\begin{tabular}[c]{@{}l@{}}Number of \\ data access \\ files\end{tabular}} & \textbf{\begin{tabular}[c]{@{}l@{}}Number of\\ refactoring\end{tabular}} \\ \midrule
Eclipse-ee4j/eclipselink & 10403 & 1371 & 43 & 80, 311 \\
Adempiere/adempiere & 15754 & 941 & 365 & 2,104,700 \\
Appirio-tech/direct-app & 3073 & 876 & 95 & 1492 \\
DotCMS/core & 17957 & 740 & 40 & 60, 211 \\
Wso2/carbon-apimgt & 33174 & 656 & 12 & 32,101 \\
Oltpbenchmark/oltpbench & 1110 & 303 & 131 & 2396 \\
Mtotschnig/MyExpenses & 9065 & 287 & 3 & 9949 \\
Querydsl/querydsl & 7874 & 249 & 23 & 38,065 \\
\begin{tabular}[c]{@{}l@{}}Wordpress-mobile/\\ WordPress-Android\end{tabular} & 59048 & 221 & 17 & 31, 647 \\
\begin{tabular}[c]{@{}l@{}}AppLozic/\\ Applozic-Android-SDK\end{tabular} & 2298 & 202 & 6 & 2136 \\
Xipki/xipki & 6328 & 193 & 21 & 88, 623 \\
Deegree/deegree3 & 8692 & 190 & 45 & 21, 459 \\ \bottomrule

\end{tabular}
}
\end{table}
\subsection{Data Processing}
Figure \ref{fig:method} provides an overview of our data extraction approach. The steps taken in the figure are labeled for easier tracking.
In this sub-section, we describe the details of the data collection and analysis approach followed to answer our research questions.
\subsubsection{Extracting List of revisions}
After we cloned each subject system, we run the \textit{git rev-list `branch`} command to get the list of revisions in the default branch of each system. This command gets commit ids of revisions from recent to oldest for the given branch. We focused our analysis on the default or master branch of each system.

\subsubsection{Extracting commit information}
One of the independent variables in this study is the commit time. To collect metadata associated with a commit including the committer time for each revision, we used PyDriller \cite{PyDriller}, a python framework for mining software repositories. This framework provides an API to collect information from a remote or locally cloned Github repository.

\subsubsection{Detecting refactoring}

There are many refactoring detection tools available (e.g., \cite{kim2010ref, dig2006automated, xing2008jdevan, silva2020refdiff,Tsantalis:TSE:2020:RefactoringMiner2.0 }). However, we used Refactoring Miner \cite{Tsantalis:TSE:2020:RefactoringMiner2.0} to detect refactoring in our subject systems for the following advantages. One, it has state-of-the-art average precision of 99.6\% and an average recall of 94\%, two, the detection approach does not require code similarity thresholds with default values obtained from empirical studies. Those thresholds affect the performance and generalization of the tools. Third, It does not require building snapshots before analysis which greatly reduces the data collection time. Refactoring Miner relies on AST-based statement matching techniques to list refactoring instances between successive revisions in a commit history. We deployed the latest version, 2.1, which supports 81 different types of refactoring instances. Refactoring Miner takes a repository and branch to analyze and generates a JSON document containing all the detected refactoring types on the history of the specified branch. 
\subsubsection{Refactoring dataset}
We merged the result of the refactoring detection for all systems and build the refactoring dataset. Each row in this dataset contains \textit{repository name, commit id, refactoring id, refactoring type, refactoring description, file name, method lines, refactoring lines}. The refactoring id is automatically generated to give a unique id for each refactoring instance; refactoring type takes one of the 81 refactoring types; refactoring description contains the description of the refactoring generated by Refactoring Miner; file name contains the list of files associated with the refactoring; method lines contains the list of method line ranges for all the methods associated with the refactoring; refactoring lines contains the list of lines touched by the refactoring. 

\subsubsection{Identifying data access refactoring instances}

We relied on import statements to identify data access classes in our subject systems. 
We particularly looked for import statements for SQL projects corresponding to the underlying persistence technologies such as Android SQLite API, JDBC, and Hibernate. The import statements include but not limited to, \texttt{android.database.sqlite}, \texttt{android.database.DatabaseUtils}, \texttt{org.hibernate.Query}, \texttt{org.hibernate.SQLQuery}, \texttt{java.sql.Statement}. This approach was used in the work of Naggy \etal \cite{nagy2018sqlinspect} who proposed the SQLInspect tool for static analysis and SQL code smell detection and on the work of Muse \etal \cite{muse2020prevalence}. To avoid the possibility of unused import statements in the code, we added another criteria for a data access class to be associated with at least one SQL query. Using this approach, we identified 18,892 refactoring instances as data access refactoring instances.

\subsubsection{Linking refactoring dataset with commit information}
To answer the second research question, we combined the refactoring dataset and the collected commit information using the commit id. The commit information contains author time and committer time. However, we used committer time for our analysis to represent the time of a system revision.

\subsubsection{Manual analysis of sample data access refactoring instances}
To answer research question \textbf{RQ3}, we manually inspected a statistically significant sample of data access refactoring instances. We have 18,892 potential data access refactoring instances. We set our sample size to 378 to achieve a 95\% confidence level and 5\% margin of error. 
Hence, we selected 378 data access refactoring instances using simple random sampling. The manual analysis aims to identify the functionalities associated with code artifacts that are targeted by the data access refactoring activities. We examined the source code, variable name, method name, and class name to identify the functionality of the code artifact. We used open coding to come up with different functional categories. 
Two people including the first author and a volunteer with four years of experience in software development and analysis labeled all the data and resolved conflicts through discussion. 

\subsubsection{Detecting SQL query and smell}
To answer \textbf{RQ4}, we run SQLInspect \cite{nagy2018sqlinspect} on our subject systems, to extract SQL queries and SQL smells. SQLInspect performs static analysis to extract SQL queries embedded in java source code and identify four types of SQL code smells namely: \textit{Implicit Columns, Fear of the Unknown, Random Selection, and Ambiguous Groups}. 

\textit{Implicit Columns} smell occurs when select queries fetch unnecessary columns from the database by using select all (*). It may cause performance issues such as bandwidth wastage and creates unnecessary coupling between the database and application code \cite{karwin2010sql}. \textit{Fear of the unknown} smell occurs when improper handling of null values and null check during data access causes unexpected error \cite{nagy2017static}. \textit{Ambiguous Groups} occurs when developers misuse the GROUP BY statement by adding columns in the select statement that are not aggregated\cite{karwin2010sql}. \textit{Random Selection} smell occurs when querying a single random row from the database which forces a full scan which has a negative performance impact for large size tables\cite{karwin2010sql}.

SQLInspect can detect queries from applications that utilize data access frameworks such as SQLite, JDBC, and Hibernate. We run SQLInspect on all snapshots that are associated with at least one data access refactoring instance. We obtain a separate dataset for query and smell instances. Each query instance is associated with commit id, class name, query value, and line number of the query location. Similarly, each smell instance is associated with commit id, class name, smell type, and location. 
\subsubsection{Linking refactoring dataset with SQL query and smell dataset}
We linked the data access refactoring dataset and the query dataset using line-level matching and method-level matching, respectively. The common criteria for both approaches is that both the refactoring and query instances must be from the same repository, the same snapshot, and belong to the same class. Line level matching is strict in the sense that the line number of the query should be one of the lines involved in the target refactoring instance. In method level matching, we match a query instance whose location is inside one of the target methods of the target refactoring instance. We used a similar approach to match the smell dataset with the refactoring dataset. While line-level matching provides a more accurate representation of association, many refactoring instances are applied to a method that contains a query and the method level matching captures the indirect association between refactoring activities and queries or smells.

\subsection{Replication Package}

We made our dataset, data collection, and data analysis scripts publicly available at \cite{replication}, to allow for replications and extensions of our work.

\section{Case study results}
This section presents the results of our four
research questions. For each research question, we present
the motivation behind the question, the analysis approach
and the findings.

\subsection{RQ1: How prevalent are refactorings in data access classes?}
\textbf{Motivation.} Since data access classes are important in data-intensive systems, studying the prevalence of refactoring activities in such classes is critical. On one hand, if the refactorings are not prevalent, it could show that they are not getting enough attention. On the other hand,  
if they are prevalent, we further investigate the characteristics of such refactorings.

\textbf{Approach.}
We use two metrics to measure the prevalence of refactorings. These metrics are:
\begin{itemize}
    \item \textit{Number of refactoring instances:} The absolute number of refactoring instances in data access or regular classes.
    
    \item \textit{Refactoring density:} defined in Equation \ref{eq:RF}. Where \textit{no of refactorings} is the number of applied refactorings and \textit{average code size} is the average number of software lines of code of the target class overall revisions. We applied Square root to reduce the scale of the code size to match the number of refactorings.

\end{itemize}

\begin{equation}
    Refactoring \  density=\frac{no \  of \  refactorings}{\sqrt{average \  code \  size}}
    \label{eq:RF}
\end{equation}

We utilize the refactoring dataset to answer this research question. In addition, we used the tool SlOCCount \footnote{https://dwheeler.com/sloccount/} to determine the code size of each snapshot of a class involved in refactoring. SLOCCount can count the physical source line of code of a java source file. We first compare the average refactoring frequency between data access and regular classes. Second, we plot the distribution of refactoring density for data access classes and regular classes using a violin plot. Finally, we examine specific types of refactoring instances and compare their prevalence.

\textbf{Findings.} Results of our analysis show that:

\textbf{$\diamond$ Refactorings are not equally prevalent in data access classes and regular classes.} 

We have \textit{18,892} refactoring instances associated with data access classes and \textit{2,454,198} refactoring instances associated with regular classes. The refactoring dataset contains \textit{1318} data access classes and \textit{81,857} regular classes. The average refactoring in data access classes (14.33) is lower than regular classes (29.98) showing that data access refactoring is less prevalent compared to regular refactoring. While the average refactoring frequency provides insight into the overall prevalence of refactoring, it gives equal weight to each class. However, there could be the case that some classes have more attention from developers compared to other classes. To investigate this, we plotted the distribution of refactoring density for data access classes and regular classes.

\textbf{$\diamond$ The refactoring density of data access classes and regular classes show a slight but statistically significant difference in distribution.}

\begin{figure}[!ht]
\centering
        \includegraphics[width=\linewidth]{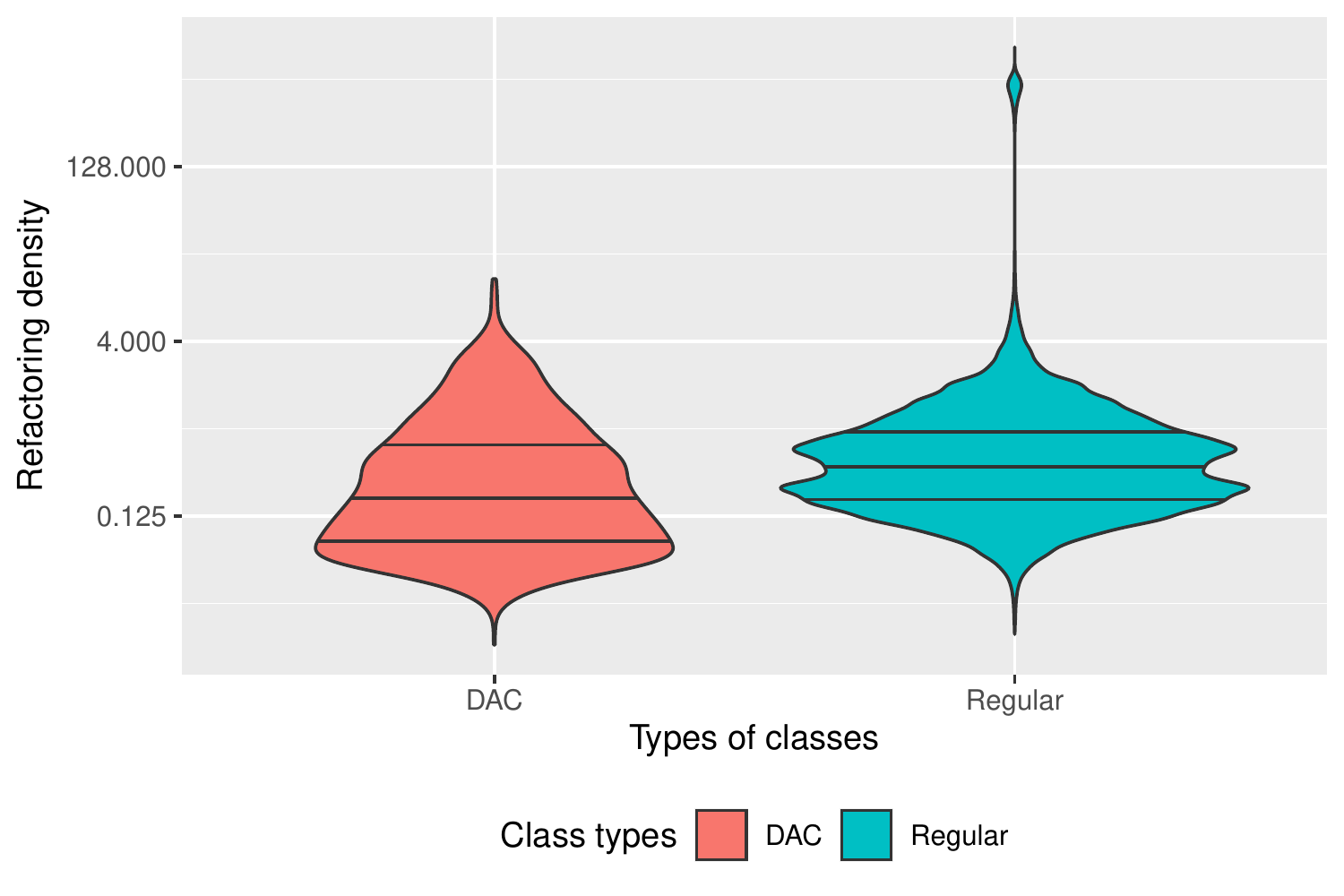}
     
    \caption{Violin Plot of the distribution of refactoring density in data access classes and regular classes}
    \label{fig:rq1-dist}
\end{figure}

Figure \ref{fig:rq1-dist} shows the distribution of refactoring density among data access classes denoted by ``DAC" and regular classes denoted by ``Regular". The 25, 50, and 75 percentiles are indicated in the plot. We can see that there is a difference in the distribution of refactoring density between data access classes 
and regular classes. 
The maximum refactoring density is 13.8 for data access classes and 1377 for regular classes. We can see that for both regular and data access classes, the refactoring density is small with median value of 0.176 and 0.335 respectively. Since the number of regular classes is much larger than the number of data access classes, we randomly selected 1318 regular classes and performed a Mann–Whitney U test to see if there is a statistically significant difference between the distribution of refactoring densities. We define the null hypothesis as \textit{$H_0$: The distribution of refactoring density between data access class and regular class is equal.}
We rejected $H_0$  
with a p-value $<$ 2.2e-16 indicating the distribution in refactoring frequencies is not equal. 
Furthermore, the 95 percent confidence interval is between 0.0823 and 0.1148, with a difference in location of 0.098 indicating that the refactoring density of data access classes is only slightly lower than that of regular classes. 

The top five data access classes in refactoring density were observed to have large code sizes and long methods. Such classes implemented core data access functions or schema definitions. The data access class with one of the highest refactoring densities 
is ApiMgtDAO.java \footnote{https://bit.ly/36XxUyk} from the project Carbon-apimgt. This class implements the API management data access functionalities with average code size of 11, 156.55 SLOC and a refactoring density of 10.34. 

\textbf{$\diamond$ The most prevalent refactoring type is \reftype{\textbf{Rename Variable}} for data access and \reftype{\textbf{Move Attribute}} for regular classes.}

We further analyzed the specific types of refactoring instances that are prevalent both in data access classes and regular classes. Table \ref{tbl:rq1-prevalence} summarizes the most prevalent refactoring types in data access and regular classes. 
The most prevalent refactoring type in data access classes is \reftype{Rename Variable} (10\%) which is associated with improving program comprehension. All the prevalent refactoring types in data access classes are either variable level or method level which shows that simpler refactoring activities that do not span more than one class are preferred by the developers over refactoring types that require touching multiple classes. Besides program comprehension, refactoring in data access classes is associated with simplifying long methods, \reftype{Extract Method} (4.7\%) is also popular refactoring activity in data access classes. 

\reftype{Move Attribute} refactoring is the most prevalent spanning 60\% of all refactoring in regular classes. \reftype{Move Attribute} refactoring is aimed at removing smells such as \textit{Shotgun Surgery} and reduces unnecessary class coupling. Another prevalent refactoring, \reftype{Change Attribute Access Modifier} (25\%), focuses on improving encapsulation. This shows that most of the refactoring on regular classes is done to improve inter-class coupling and encapsulation.

\begin{table*}[]
\centering
\caption{Top ten most prevalent refactoring types. The table shows the number of refactoring instances (count) and percentage against the total number of data access and regular class refactoring instances.}
\label{tbl:rq1-prevalence}
\resizebox{0.7\textwidth}{!}{%
\begin{tabular}{@{}lll|lll@{}}
\toprule
\multicolumn{3}{c}{\textbf{Data access class}} & \multicolumn{3}{c}{\textbf{Regular class}}          \\ \midrule
\textbf{Refactoring Type} & \textbf{count} & \textbf{percentage} & \textbf{Refactoring Type} & \textbf{count} & \textbf{percentage} \\ \midrule
\textbf{Rename Variable}                & 1880  & 9.951 & \textbf{Move Attribute}                   & 1470365 & 59.912 \\
Change Variable Type           & 1539  & 8.146 & Change Attribute Access Modifier & 600345  & 24.462 \\
Add Parameter                  & 1119  & 5.923 & Add Method Annotation            & 26913   & 1.097  \\
Add Parameter Modifier         & 998   & 5.283 & Rename Method                    & 22963   & 0.936  \\
Change Parameter Type          & 910   & 4.817 & Change Parameter Type            & 21618   & 0.881  \\
Rename Parameter               & 888   & 4.7   & Change Variable Type             & 21372   & 0.871  \\
Extract Method                 & 887   & 4.695 & Change Return Type               & 19459   & 0.793  \\
Add Method Annotation          & 816   & 4.319 & Add Parameter                    & 18528   & 0.755  \\
Rename Method                  & 705   & 3.732 & Rename Variable                  & 18345   & 0.747  \\
Change Method Access Modifier  & 620   & 3.282 & Rename Parameter                 & 16107   & 0.656  \\ \bottomrule
\end{tabular}%
}
\end{table*}

\begin{tcolorbox}[colback=white, colframe=black,left=2pt,right=2pt,top=3pt,bottom=1pt]
\textbf{RQ1 Summary:}
Refactorings are slightly less prevalent in data access classes compared to regular classes. Furthermore, \reftype{\textbf{Rename Variable}} is the most prevalent refactoring type in data access classes. Data access refactoring activities are concentrated on few core classes.  
\end{tcolorbox}

\subsection{RQ2: How do refactoring activities change during the lifetime of the subject systems?}

\textbf{Motivation.} 
Software becomes complex as it evolves due to the added and improved features. It is interesting to study if the evolution of software determines the type and prevalence of refactoring activities performed by developers on data access classes. If the refactoring activities are equally frequent in all stages, it shows that developers considered refactoring as a regular activity. Otherwise, it may suggest that refactoring activities in data-intensive systems are triggered by the increasing complexity introduced during evolution.   

\textbf{Approach.} We use \textit{relative commit time} as a metric to express when a refactoring happens. Since every refactoring is associated with a commit, we use the associated commit time. Due to the variation in the maturity of the subject systems we use relative time rather than an absolute time for correct comparison. \textit{Relative commit time} is computed using Equation \ref{eq:relcommittime} where distance is computed as the number of commits the subject system has at the time of the refactoring and \textit{totalCommits} is the total number of commits the subject system has at the time of the experiment.      
\begin{equation}
\label{eq:relcommittime}
  \textit{RelativeCommitTime} (\%)= \frac{distance*100}{TotalCommits}
\end{equation}

We linked the commit information and the refactoring dataset using the commit id. Then, we computed the  \textit{relative commit time} for each refactoring using Equation \ref{eq:relcommittime}. We first show the distribution of \textit{relative commit time} for data access refactoring and regular refactoring instances using violin plots. Second, we statistically compare the two distributions using Mann–Whitney U test. Third, we compare the relative commit time between data access and regular refactorings at the subject system level. Finally, we show the summary of the relative commit time distribution for most prevalent data access refactoring types. 

\textbf{Findings.} Results show that: 

\textbf{$\diamond$ The median relative commit time for data access refactorings is higher compared to regular refactorings.}

 \begin{figure}[!ht]
\centering
        \includegraphics[width=\linewidth]{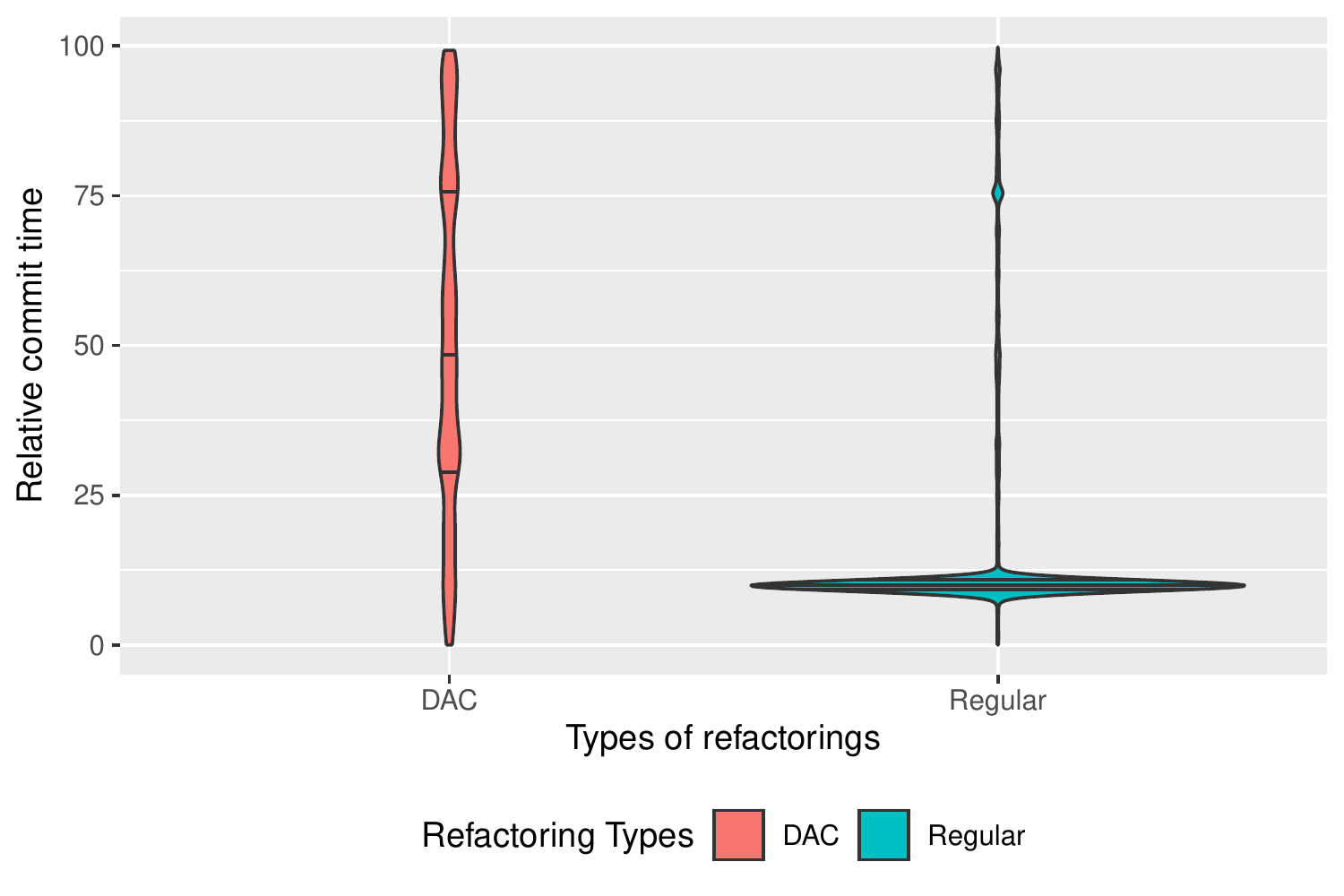}
     
    \caption{Violin Plot of the distribution of relative commit time in data access refactorings (DAC) and regular refactorings}
    \label{fig:rq2-dist-reg-dac}
\end{figure}

Figure \ref{fig:rq2-dist-reg-dac} shows the distribution of relative commit time for data access refactorings and regular refactorings. As can be seen on Figure \ref{fig:rq2-dist-reg-dac}, the relative commit time is different between data access and regular refactorings. A large number of regular refactorings occurred in the first 25\% relative commit time or during the beginning stages of the projects. However, for data access classes, the refactoring activities are distributed across all project evolution stages with more tendency to later stages. The median relative commit time for data access refactorings (48.35\%) is higher compared to regular counterparts (16.88\%). Furthermore, 25\% of data access refactorings occurred below relative commit time of 30.31\% while 25\% of regular refactorings occurred below 9.89\%. This shows that developers often perform regular refactorings during the beginning stages of the subject systems evolution while they perform data access refactorings throughout the evolution of the subject systems.

\textbf{$\diamond$ The difference in the distribution of relative commit time between data access and regular refactorings is statistically significant.}

We performed Mann–Whitney U test on relative commit time using all data access refactorings (18886) and equal size sample of regular refactorings, and rejected the null hypothesis with (W = 6458096, p-value $<$ 2.2e-16). This indicates that the difference in distribution between the relative commit time of data access refactorings and regular refactorings is statistically significant.

\textbf{$\diamond$ Data access refactorings have a higher median relative commit time in the majority of the subject systems.}

We further analyzed the relative commit time between data access refactorings and regular refactorings by splitting the data by subject systems. The median relative commit time is higher for data access refactorings compared to regular refactorings in 7 out of 12 subject systems (58.3\%). Four of the five systems where the regular refactoring commit time is higher are associated with a low number of data access classes and data access refactorings. The low number of data access refactorings is expected given the low number of data access classes. 

\textbf{$\diamond$ The \reftype{Add Method Annotation} refactoring often occur at later stages of the evolution of the subject systems and in data access classes.}

Another interesting analysis would be to investigate what type of data access refactorings are performed by developers at different stages of the subject systems evolution. We focused our analysis on the top 10 most prevalent data access refactoring types from \textbf{RQ1}. Table \ref{tbl:rq2-rel-commit-distribution} shows the summary of the distribution of relative commit time for the top ten most prevalent data access refactorings sorted by the median. If we divide the median of relative commit time into four quartiles, we can see that \reftype{Add Method Annotation} belongs to the fourth quartile indicating that this refactoring is often performed by developers at the latest stage of the evolution of the systems. The median relative commit time of the other refactoring types are found at the second and third quartiles, indicating that they are performed in the middle stages of the evolution of the subject systems. On the other hand, \reftype{Add Parameter Modifier} is often performed at the early stages (median=33.668\%) of the evolution of the subject systems with 75\% of such refactorings having relative commit time below 33.683\%.


\begin{table*}[!ht]

\centering
\caption{Distribution of relative commit time for the top ten prevalent data access refactoring types.}
\label{tbl:rq2-rel-commit-distribution}
\resizebox{0.7\textwidth}{!}{%
\begin{tabular}{@{}lllllllll@{}}
\toprule
\textbf{Refactoring Type} & \textbf{count} & \textbf{mean} & \textbf{std} & \textbf{min} & \textbf{25\%} & \textbf{50\%} & \textbf{75\%} & \textbf{max} \\ \midrule
Add Method Annotation & 816 & 69.037 & 30.263 & 4.248 & 44.532 & \textbf{85.470} & 93.424 & 98.707 \\
Rename Parameter & 888 & 56.270 & 30.873 & 0.024 & 30.174 & \textbf{57.387} & 82.016 & 99.201 \\
Rename Variable & 1880 & 54.250 & 25.696 & 0.051 & 34.077 & \textbf{56.409} & 76.458 & 99.243 \\
Extract Method & 887 & 52.945 & 29.327 & 0.245 & 28.157 & \textbf{53.866} & 77.619 & 99.193 \\
Change Parameter Type & 910 & 52.266 & 30.011 & 0.051 & 29.075 & \textbf{50.170} & 81.293 & 98.994 \\
Rename Method & 705 & 50.903 & 27.934 & 0.391 & 30.028 & \textbf{48.467} & 74.205 & 99.201 \\
Change Method Access Modifier & 620 & 51.837 & 30.693 & 0.214 & 21.556 & \textbf{48.467} & 83.764 & 99.092 \\
Change Variable Type & 1539 & 49.893 & 28.986 & 0.265 & 27.889 & \textbf{48.139} & 78.490 & 98.839 \\
Add Parameter & 1119 & 44.238 & 29.789 & 0.051 & 20.219 & \textbf{39.075} & 71.615 & 99.066 \\
Add Parameter Modifier & 998 & 37.624 & 13.028 & 9.175 & 33.652 & \textbf{33.668} & 33.683 & 98.016 \\ \bottomrule
\end{tabular}
}
\end{table*}

\begin{tcolorbox}[colback=white, colframe=black,left=2pt,right=2pt,top=3pt,bottom=1pt]
\textbf{RQ2 Summary:}
 The median relative commit time for data access refactorings (48.35\%) is higher than that of the regular refactorings (16.88\%), indicating that developers have a tendency to refactor data access classes at later stages of the evolution of systems compared to regular classes. Among the most prevalent data access refactoring types, \reftype{Add  Method  Annotation} has a tendency to be applied at the later stages of the software evolution (median=85.47\%). 
\end{tcolorbox}

\subsection{RQ3: What do code elements targeted by data access refactorings implement?}

\textbf{Motivation.}
In \textbf{RQ1}, we have shown that refactoring activities are prevalent in data access classes. However, not all components of data access classes are directly associated with data access. It is expected that data access classes could also contain constructors, accessors, mutators, and non data access logic implementations. Hence, it is necessary to investigate if data access refactorings focus on the actual data access logic or other non data access functionalities.

\textbf{Approach.}
To identify functionalities of code artifacts associated with refactoring in data access classes, we randomly sampled 378 data access refactoring instances and manually analyzed the code associated with each refactoring. Out of the 378 analyzed samples, 10 were false positives. In addition, We were not able to assign functionality to 2 instances since they were associated with empty methods and their method name is not descriptive enough. Finally, we remained with 366 refactoring instances. We first describe each of the identified functionalities and then discuss the prevalence of the functionalities using absolute numbers and percentages against the total sample.

\textbf{Finding.} \textbf{Fetching data and inserting data were the most prevalent functionalities implemented by refactored code.}

\begin{figure}[ht]
\centering
        \includegraphics[width=\linewidth]{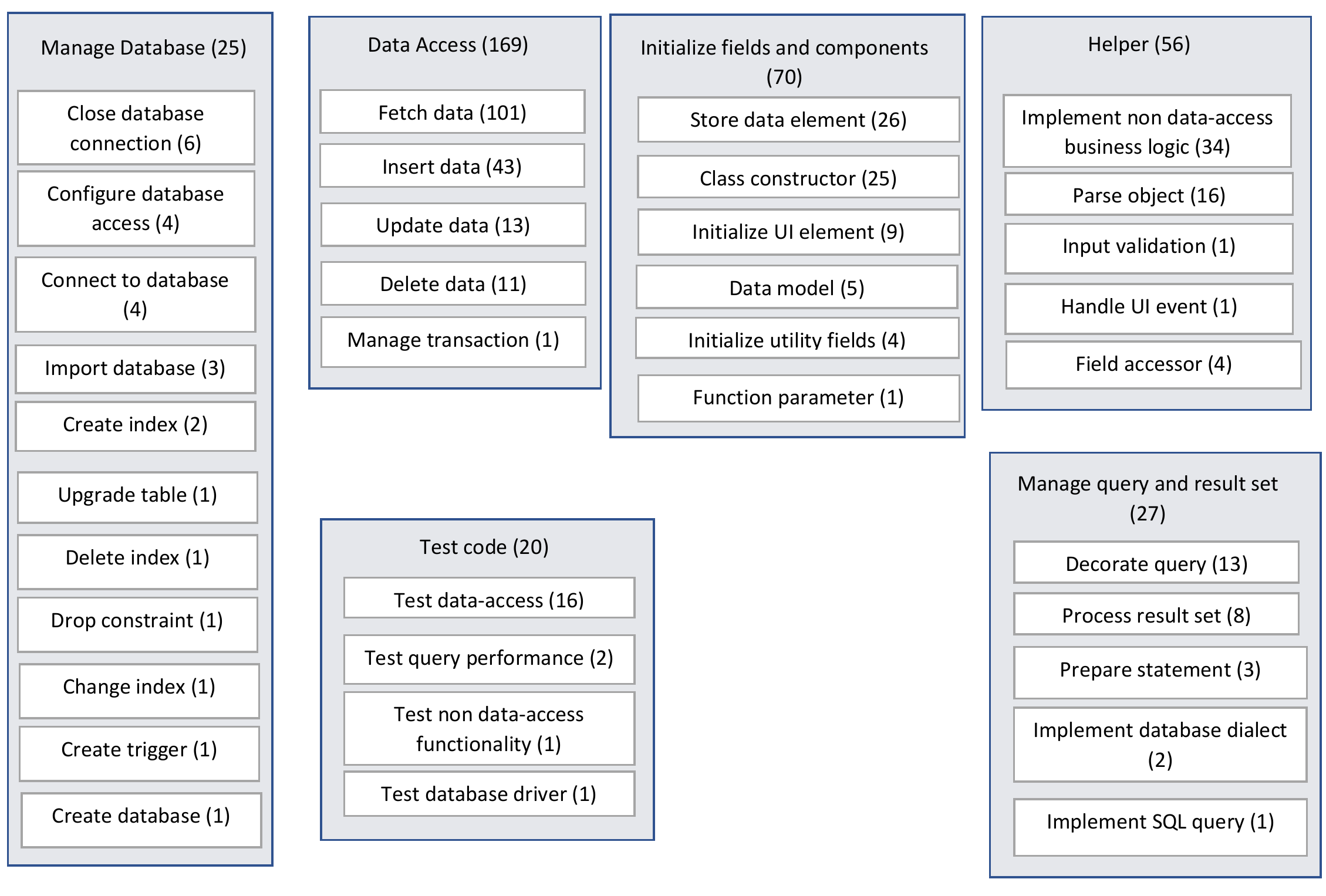}
     
    \caption{Functionalities of code artifacts associated with refactoring in data access classes. The sub-categories are ordered from the most prevalent to the least prevalent.}
    \label{fig:rq3-cat}
\end{figure}

  Figure \ref{fig:rq3-cat} shows the identified functionalities categorized by seven higher-level categories. The number of refactoring instances is indicated for the categories as well as each functionality. We will describe the categories in the following paragraphs.

$\diamond$ \textit{Data Access:}
Refactoring instances that target code elements that perform read and write operations on the data stored in databases are categorized under this category. We categorized code elements that contain one or more data manipulation SQL queries (SELECT, INSERT, UPDATE and DELETE) and code elements that are associated with managing transactions under data access. A large number of refactoring instances (46.17\%) are associated with data access. In particular, \textit{Fetch data}, a functionality associated with reading some data from database is the most prevalent (27.6\%) followed by \textit{Insert data} (11.74\%) and \textit{Update data} (3.55\%). One refactoring instance was associated with \textit{Manage Transaction} and more specifically rolling back a transaction.

$\diamond$ \textit{Initialize fields and components:}
Refactoring instances that target code elements that initialize or set data fields, utility fields, or user interface (UI) components are categorized under this functionality. Data fields are variables that represent a database entity and whose values are used to capture data from a database or to populate a database. On the other hand, utility fields are used to capture non-data access related entities. The \textit{Initialize fields and components} functionality is associated with 19.12\% of the refactoring instances. The \textit{Store data element} functionality associated with initializing data fields is the most prevalent functionality associated with 7.1\% refactoring instances, in this group followed by \textit{Class constructor} (6.83\%) and \textit{Initialize UI element} (2.45\%), associated with initializing graphical user interface components with data obtained from a database or as a result of some intermediate transformation of data. \textit{Data model}, a functionality associated with providing an abstraction of database entities is associated with 1.36\% of the refactoring instances. The remaining functionalities i.e., \textit{Initialize utility fields, Field accessor (get the value of a field) and function parameter (storing passed value as function parameters) 
make 2.46\% of refactoring instances.}

$\diamond$ \textit{Helper:}
This category regroups refactoring instances targeting code elements that are involved in the implementation of business logic that is not directly associated with database access, such as parsing objects (access elements of complex objects), validation of user input, and handling UI events. The \textit{Helper} functionality is associated with 15.03\% of refactoring instances. \textit{Implement non data access business logic} is the most prevalent functionality in this group. It is associated with 9.3\% refactoring instances followed by \textit{Parse object} which is associated with 4.1\% instances. \textit{Input validation} and \textit{Handle UI event} functionalities are less prevalent; each are associated with only 1 instance.

$\diamond$ \textit{Manage query and result set}: Refactoring instances that target code elements associated with manipulating the query, parsing, and transforming the query result set are grouped under this category. This group accounts for 7.38\% of refactoring instances. Specifically, \textit{Decorate query}, functionality associated with pre-processing a query before passing to the database takes the largest share (3.55\%) followed by \textit{Process result set} (2.19\%), associated with parsing and iterating over query results. \textit{Prepare statement}, associated with parameterizing queries, \textit{Implement database dialect}, associated with defining particular features of the SQL query available when accessing a data entity, and \textit{Implement SQL query}, associated with abstracting SQL data access over data sources that do not directly support SQL queries, account for 1.64\% in total.    

$\diamond$ \textit{Manage database}: We categorize refactoring instances that target code elements involved in connecting with database, managing the index, managing constraints, managing triggers, and importing the database under this group. The \textit{Manage database} functionality is associated with 6.83\% refactoring instances. \textit{Close database connection} functionality, associated with terminating a database connection and releasing resources, is the most prevalent in this group, followed by \textit{Configure database access} (1.09\%), associated with setting the connection parameters and credentials, and \textit{Connect to database} (1.09\%). The \textit{Import database} functionality is associated with creating a database with a template or data obtained from external files and accounts for 0.82\% of the refactoring instances. \textit{Create index, Delete index, change index} functionalities are focused on managing indexes and account for 1.09\% of the refactoring instances. The remaining functionalities (\textit{Upgrade table, drop constraint, create trigger and create database} are associated with one refactoring instance each.) 

\vspace{-1.1pt}
$\diamond$ \textit{Test code}: Refactoring instances that target code elements involved in testing production code elements are categorized under \textit{Test code}. The \textit{Test code} functionality is associated with 5.46\% of the refactoring instances. The most common functionality in \textit{Test code} is testing production code involved in database read and write operations, known as \textit{Test data access}. It accounts for 4.37\% of refactoring instances. On the other hand, the \textit{Test non data access functionality}, associated with testing production code that is not directly associated with database access is associated with 1 refactoring instance. Code elements involved with testing the performance of a certain query are categorized as \textit{Test query performance} and they are associated with only two refactoring instances. Code elements that specifically test database drivers are categorized as \textit{Test database driver}. 

\begin{tcolorbox}[colback=white, colframe=black,left=2pt,right=2pt,top=3pt,bottom=1pt]
\textbf{RQ3 Summary:}
  Data access functionality is associated with a large number of refactoring instances accounting for 46.17\% of the analyzed data access refactorings. Among data access functionalities, fetching data takes the largest share accounting for 27.6\% of the analyzed data access refactorings followed by inserting data.
\end{tcolorbox}

\subsection{RQ4: Do data access refactoring activities touch SQL queries and SQL code smells? }

\textbf{Motivation.}
In \textbf{RQ3}, we showed that fetching data and inserting data
are associated with most of the data access refactoring instances. All our subject systems use SQL for database interactions including fetching and inserting data. Hence, it is interesting to investigate if SQL statements change during data access refactoring activities. Furthermore, SQL code smells are shown to be prevalent in data-intensive systems \cite{muse2020prevalence} and it is also interesting to see if queries that contain such smells are touched during data access refactorings.  

\textbf{Approach.}

To investigate if data access refactoring instances touch SQL queries and SQL code smells, we matched the refactoring dataset with query dataset and smell dataset using \textit{line level} and \textit{method level} matching. \textit{Line level} matching has more 
strict criteria than method level matching since the line number of the query or smell should be the same as the line number of the code involved in refactoring. On the other hand, \textit{Method level} matching is less restrictive as it checks if the query or smell is part of the method involved with the refactoring. For both cases, the subject system, revision, and class Name should match between the query, smell, and refactoring dataset. 

\textbf{Findings.} Results show that:

\textbf{$\diamond$ 
Only small fractions of data access refactorings touch code lines containing SQL query.}

\subsubsection{Data access refactoring and SQL queries}
Our result shows that a small number of refactoring instances touched SQL queries (using line level matching). We have 18, 892 data access refactoring instances out of which only 86 instances (0.45\%) touched an SQL query. When we further analyze the types of the 86 refactoring instances involved with SQL queries, the most prevalent type is \textit{Change Return Type} with 17 refactoring instances (19.76\%), followed by \textit{Add Parameter} with 12 refactoring instances (13.95\%), \textit{Extract Method} with 10 refactoring instances (11.63\%), and \textit{Change Variable Type} with 7 refactoring instances (8.14\%).

\textbf{}{$\diamond$ 30\% of data access refactorings were applied on a method that contains SQL queries.}

When we consider method level matching, we get more SQL queries associated with the refactoring instances as expected. Results show that 5607 refactoring instances (29.68\%) contained SQL queries inside the target methods. When we see the refactoring types, the most prevalent type is \textit{Rename Variable} with 821 instances (14.64\%), followed by \textit{Change Variable Type} with 603 instances (10.75\%), \textit{Add Parameter} with 409 instances (7.29\%), and \textit{Change Parameter Type} with 355 instances (6.33\%). 
Such types of refactorings focus on fixing lexical smells or reflecting API changes and are not associated with modifying query or improving SQL code smells.


\subsubsection {Data access refactoring and SQL smells}

 SQL smells are detected from SQL queries. Hence, the smell dataset is a subset of the query dataset. Although SQLInspect can detect \textit{Implicit columns, Fear of the Unknown, Ambiguous Groups and Random Selection}, the smell dataset only contains instances of \textit{Implicit columns} smell and \textit{Fear of the unknown} SQL smell. Indeed both types of SQL smells are shown to be more prevalent in data-intensive systems \cite{muse2020prevalence}. 

\textbf{$\diamond$ 1.35\% of data access refactorings were involved with a method containing queries with a SQL code smell.}

Using the line level matching, we did not find any instance of SQL code smell associated with data access refactoring instances. On the other hand, we find 256 refactoring instances (1.355\%) whose target method contains a query with SQL Code smell. From the 256 refactoring instances, \textit{Rename Variable} takes the larger share with 60 instances (23.44\%), followed by \textit{Extract Method} with 30 instances (11.72\%), \textit{Rename Method} with 26 instances (10.156\%), \textit{Add Parameter} with 24 instances (9.38\%), and \textit{Change Variable Type} with 21 instances (8.2\%).

When examining the specific types of SQL code smells associated with the refactoring instances, we observe that 200 instances (78.12\%) are associated with only \textit{Implicit Columns} SQL smell, and 49 instances (19.14\%) are associated with only the \textit{Fear of the unknown} SQL smell.

\begin{tcolorbox}[colback=white, colframe=black,left=2pt,right=2pt,top=3pt,bottom=1pt]
\textbf{RQ4 Summary:}
  We observed that only a small fraction of data access refactoring instances touched SQL queries and SQL smells. Using line level matching, only 0.45\% of refactoring instances were involved with SQL queries. Furthermore, We did not find any SQL code smell instance associated with refactoring. Using a method level matching, we found 29.68\% refactoring instances associated with SQL queries and 1.35\% refactoring instances associated with SQL code smells.  
\end{tcolorbox}
\section{Implication of findings}

We believe that our findings contribute towards characterizing refactoring activities in data access classes. Our findings show that data access refactorings have a small prevalence considering their importance to the subject systems. Furthermore, the prevalent refactoring types are mostly simple refactorings aiming at improving the code understandability. These findings show that further investigation is needed to identify why data access refactorings do not focus on improving the data access performance, by for example fixing SQL code smells, but rather focus on low-level refactorings like the ones performed on non data access classes. 

Many reasons such as lack of developers' awareness of the impacts of SQL code smells and lack of refactoring tool support could explain the lack of focus on improving data access performance by refactoring. Hence, we recommend that the research community examines this issues and work on proposing and--or improving refactoring recommendation tools.

Our findings regarding the functionalities implemented by data access classes that are often targeted by refactoring could help in the identification of data access specific refactoring types (Eg. refactoring query statements) and the subsequent development of refactoring recommendation tools that are aware of data access specific refactoring types by guiding on what part of data access operations to focus research and development effort.

Our findings help practitioners such as developers and quality assurance teams by providing insights on what aspects of data access operation issues are addressed by current refactoring efforts and what aspects are not addressed by refactoring. This could be leveraged by developers and the quality assurance team on prioritizing refactoring activities and defining the standards and objectives to be achieved during refactoring of data access classes.  

\section{Threats to validity}
\subsection{Threats to construct validity}
We relied on state-of-the-art refactoring detection and SQL code smell detection tools to extract refactoring instances, SQL queries, and SQL code smells. Refactoring Miner is reported to achieve 99\% precision and 94\% recall. However, we could still miss some refactoring instances and the tool might introduce false positives. The SQL Inspect tool used to identify SQL queries and SQL code smells could also miss some queries and smells. The interpretation of our findings should take this into account. Another potential threat to construct validity comes from the potential researcher bias in the manual analysis of \textbf{RQ3}. To overcome this threat, the manual analysis conducted by the first author was evaluated by an independent researcher and all the disagreements in labeling were resolved with discussion.

\subsection{Threats to internal validity} We did not claim any causation as this is an exploratory study. This study does not have threats to internal validity.

\subsection{Threats to conclusion validity} Conclusion threat to validity could be associated with the statistical analysis approach. We only used non-parametric statistical tests. Hence, our study does not have conclusion threats to validity.

\subsection{Threats to external validity}

Since we performed a longitudinal study, we have to limit our subject systems to 12. This could limit the external validity of our study. However, we carefully selected our subject systems to represent open source data-intensive systems by considering the number of SQL queries as a proxy. Furthermore, those systems come from different application domains and rely on different data access technologies including JDBC, SQLite, and Hibernate. Hence, our findings can be generalized to the extent of open source data-intensive systems.

\subsection{Threats to reliability validity} To minimize potential threats to reliability, all our subject systems are open source and available on GitHub. Furthermore, we provided all the necessary materials to replicate our study in our replication package \cite{replication}. 
\section{Related work}
In this section, we provide an overview of the state of the art in refactoring detection, empirical study on refactoring, and empirical studies in data-intensive systems.

\subsubsection{Empirical studies on refactoring} Many empirical studies explored the prevalence, nature, co-occurrence, and impact of refactoring activities on software quality. Since Fowler proposed a catalog of refactoring types \cite{fowler2002refactoring}, there have been many studies on refactoring activities. In the following, we will focus on the most recent studies for brevity. Silva \etal \cite{silva2016we}, surveyed open source developers to identify the motivations behind applying refactoring and found that refactorings are not motivated by code smells. They are rather motivated by changes in software requirements such as bug fixing and feature enhancement. Chavez \etal\cite{chavez2017does} studied the impact of refactoring activities on internal quality attributes such as cohesion, coupling, complexity, and inheritance and found that 65\% of refactoring instances improved the associated internal quality attributes. Ferreira \etal\cite{ferreira2018buggy} analyzed 20,689 refactoring instances from 5 open-source projects to study the relationship between refactoring activity and bugs and found that code elements involved in floss refactoring are more bug-prone compared to root canal refactoring. Mahmoudi \etal\cite{mahmoudi2019refactorings} conducted an empirical study to investigate the impact of refactoring activities on merge conflicts using 3000 java subject systems and found that 22\% of the refactoring instances were involved with merge conflicts. Vassallo \etal\cite{vassallo2019large} studied 200 open source projects belonging to different java ecosystems and showed that the type of refactoring operations applied by developers depends on the support of development environments. Furthermore, they showed that planning for refactoring activities is done based on the age of the software component and proximity to software release. Peruma \etal\cite{peruma2019preliminary} explored refactoring activities in android applications and found that \textit{rename attribute} is the most common refactoring in android applications. They also found that the overall motivation of refactoring is quality improvement exploiting refactoring commit messages. 
Iammarino\etal\cite{iammarino2019self} studied the co-occurrence of refactoring activities and SATD removals using a curated SATD dataset and the Refactoring Miner tool and found that refactorings are more likely to co-occur with SATD removal commits than with other commits, however, in most cases, they belong to different quality improvement activities rather than part of the SATD removal. Rename refactorings are specifically studied given the importance of such refactorings on program comprehension. Peruma\etal\cite{peruma2018empirical} analyzed 524,113 rename refactorings and found that most rename refactorings narrow the meaning of the identifiers for which they are applied. In another study, Peruma\etal\cite{peruma2019contextualizing} found that rename refactorings are more preferred by less experienced developers and that developers frequently change the semantic meaning after a rename refactoring.

\subsubsection{Refactoring detection approaches} The aforementioned empirical studies were possible thanks to the introduction of different refactoring detection approaches based on: object-oriented metrics\cite{demeyer2000finding},vector space information retrieval\cite{antoniol2004automatic}, clone detection\cite{weissgerber2006identifying}, syntactic and semantic analysis of source code\cite{dig2006automated}, UMLDiff algorithm\cite{xing2008jdevan}, and template logic query\cite{kim2010ref}. Silva\etal\cite{silva2020refdiff} proposed RefDiff 2.0, a multi-language refactoring detection tool utilizing code structure tree. In our study, we used Refactoring Miner proposed by Tsantalis\etal\cite{Tsantalis:TSE:2020:RefactoringMiner2.0} due to its state of the art detection accuracy, faster execution time, and because it does not require similarity thresholds as an input.
\subsubsection{Study on SQL code smells}
Data access operations play a pivotal role in data-intensive systems. The implementation of data access operations could suffer from data access specific code smells such as SQL code smells besides traditional code smells. SQL being primarily used as a data access language in data-intensive systems, anti-patterns in SQL could adversely affect the performance and maintainability of such systems in the long run. The book of Karwin \cite{karwin2010sql} provided a comprehensive catalog of  SQL anti-patterns. Since the introduction of SQL anti-patterns, few automatic detection approaches and tools have been proposed. Khumnin et al. \cite{Khumnin2017} provided a tool for detecting logical database design anti-patterns from Transact-SQL queries. Another tool, Sharma \etal \cite{Sharma2018} provided DbDeo for the detection of \emph{database schema} smells. Furthermore, Nagy and Cleve \cite{nagy2017static} proposed a SQL code smell detection approach using static analysis to extract SQL queries embedded in java code and to detect SQL code smells. They proposed the SQLInspect \cite{nagy2018sqlinspect} tool based on this approach. There are also several SQL analysis and smell detection tools such as SQL Prompt\footnote{\url{https://www.red-gate.com/hub/product-learning/sql-prompt}} and SQL Enlight\footnote{\url{https://sqlenlight.com/}}. However, such tools require users to provide SQL queries and therefore can't analyze queries embedded in source code. Hence, we relied on SQLInspect to detect SQL code smells in our subject systems.

There are very few empirical studies on SQL code smells.
 De Almeida Filho \etal\cite{de2019prevalence} investigate the prevalence and co-occurrence of SQL code smells in PL/SQL projects. Arzamasova\etal\cite{Arzamasova2018}. proposed a detection approach to capture anti-patterns in SQL logs. They were able to detect six anti-patterns by analyzing skyServer log dataset. Muse\etal\cite{muse2020prevalence} studied the prevalence and impact of SQL code smells on 150 open source java projects. They showed that SQL code smells are prevalent in data-intensive systems but they have a weak co-occurrence with other code smells and with bugs. Furthermore, they provided evidence indicating that SQL code smells are persistent across the evolution of subject systems. Shao \etal provided a catalog of performance anti-patterns in database-backed web applications\cite{shao2020database}.

\section{Conclusion}

In this study, we investigated the prevalence, evolution, functionality, and co-occurrence of refactoring activities with data access code smells by conducting a longitudinal study on 12 open source data-intensive projects. We analyzed 174,776 commits and extracted 2,473,090 refactoring instances to build the refactoring dataset. We also extracted SQL queries and SQL data access smells from the same subject systems and linked them with the refactoring dataset. First, we compared the type of refactoring activities between data access classes and regular classes to investigate if developers perform different types of refactoring between them. Second, we computed the distribution of refactoring instances over the evolution of the subject systems to investigate if the refactoring applied by developers vary between data access classes and regular classes and varies across systems evolution. Third, we conducted a manual analysis of over 378 sample data access refactoring instances to identify the functionality of their target codes. Finally, we investigated if data access refactoring activities touch SQL queries and SQL smells.

Our results show that different types of refactoring instances are prevalent in data access classes and regular classes. The most prevalent refactoring is \textit{Rename variable} in data access classes and \textit{Move attribute} in regular classes, respectively. Furthermore, data access refactoring tends to be applied at later stages of the subject systems evolution compared to non data access refactoring. The result of the manual analysis showed that a large number of data access refactoring instances target code that implements data access in general and data fetching in particular. We also find that data access refactoring instances do not generally touch both SQL queries and SQL code smells.

Overall our results show that data access refactorings although prevalent, do not directly target statements that directly interact with databases (e.g., SQL statements). This calls for future research on refactoring detection of data access specific smells and on the characteristics of such data access smell refactorings. 
or  do to the lack of tool support to perform complex refactorings. We plan to examine this hypothesis as a future work.

\textbf{Acknowledgement:}
This work is partly funded by the Natural Sciences and Engineering Research Council of Canada (NSERC) and the Fonds de Recherche du Québec (FRQ).






%
\bibliographystyle{IEEEtran}
\balance
\bibliography{citation}

\end{document}